\documentclass[%
reprint,
superscriptaddress,
amsmath,amssymb,
prb,
longbibliography
]{revtex4-2}

\usepackage{color}
\usepackage{xcolor}
\usepackage{graphicx}% Include figure files
\usepackage{dcolumn}% Align table columns on decimal point
\usepackage{siunitx}
\usepackage{bm}% bold math
\usepackage{hyperref}% add hypertext capabilities
\hypersetup{colorlinks=true,linkcolor=blue,citecolor=blue,urlcolor=blue}
\usepackage{soul} %\st delete, \hl, highlight, \ul, _, \caps
\usepackage[colorinlistoftodos]{todonotes}

\begin{document}

\preprint{FeGe phase diagram}

\title{Annealing-tunable charge density wave in the kagome antiferromagnet FeGe}
%Nature of the annealing tunable charge density wave in the kagome antiferromagnet FeGe
%Annealing tunable charge density wave in the magnetic kagome material FeGe
%Correlation between structure, magnetism, and charge density wave in the kagome magnet FeGe revealed by annealing effect\textcolor{blue}{s}
%Nature of charge density wave in the kagome antiferromagnet FeGe revealed by annealing effects

\author{Xueliang Wu}
\thanks{These authors contributed equally to this work}
\affiliation{Low Temperature Physics Laboratory, College of Physics and Center of Quantum Materials and Devices, Chongqing University, Chongqing 401331, China.
}

\author{Xinrun Mi}
\thanks{These authors contributed equally to this work}
\affiliation{Low Temperature Physics Laboratory, College of Physics and Center of Quantum Materials and Devices, Chongqing University, Chongqing 401331, China.
}

\author{Long Zhang}
\thanks{These authors contributed equally to this work}
\affiliation{Low Temperature Physics Laboratory, College of Physics and Center of Quantum Materials and Devices, Chongqing University, Chongqing 401331, China.
}

\author{Chin-Wei Wang}
\affiliation{National Synchrotron Radiation Research Center, Hsinchu 30077, Taiwan.
}

\author{Nour Maraytta}
\affiliation{Institute for Quantum Materials and Technologies, Karlsruhe Institute of Technology, Kaiserstraße 12, 76131 Karlsruhe, Germany.}

\author{Xiaoyuan Zhou}
\affiliation{Low Temperature Physics Laboratory, College of Physics and Center of Quantum Materials and Devices, Chongqing University, Chongqing 401331, China.
}

\author{Mingquan He}
\email{mingquan.he@cqu.edu.cn}
\affiliation{Low Temperature Physics Laboratory, College of Physics and Center of Quantum Materials and Devices, Chongqing University, Chongqing 401331, China.
}

\author{Michael Merz}
\email{michael.merz@kit.edu}
\affiliation{Institute for Quantum Materials and Technologies, Karlsruhe Institute of Technology, Kaiserstraße 12, 76131 Karlsruhe, Germany.}
\affiliation{Karlsruhe Nano Micro Facility, Karlsruhe Institute of Technology, Kaiserstraße 12, 76131 Karlsruhe, Germany.}

\author{Yisheng Chai}
\email{yschai@cqu.edu.cn}
\affiliation{Low Temperature Physics Laboratory, College of Physics and Center of Quantum Materials and Devices, Chongqing University, Chongqing 401331, China.
}

\author{Aifeng Wang}
\email{afwang@cqu.edu.cn}
\affiliation{Low Temperature Physics Laboratory, College of Physics and Center of Quantum Materials and Devices, Chongqing University, Chongqing 401331, China.
}

\date{\today}

\begin{abstract}
The unprecedented phenomenon that a charge density wave (CDW) emerges inside the antiferromagnetic (AFM) phase indicates an unusual CDW mechanism associated with magnetism in FeGe.  Here, we demonstrate that both the CDW and magnetism of FeGe can be effectively tuned through post-growth annealing treatments. Instead of the short-range CDW reported earlier, a long-range CDW order is realized below 110 K in single crystals annealed at \SI{320}{\degreeCelsius} for over 48 h. The CDW and AFM transition temperatures appear to be inversely correlated with each other. The onset of the CDW phase significantly reduces the critical field of the spin-flop transition, whereas the CDW transition remains stable against minor variations in magnetic orders such as annealing-induced magnetic clusters and spin-canting transitions. Single-crystal x-ray diffraction measurements reveal substantial disorder on the Ge1 site, which is characterized by displacement of the Ge1 atom from the Fe$_3$Ge layer along the $c$ axis and can be reversibly modified by the annealing process. The observed annealing-tunable CDW and magnetic orders can be well understood in terms of disorder on the Ge1 site. Our study provides a vital starting point for the exploration of the unconventional CDW mechanism in FeGe and of kagome materials in general. 
\end{abstract}
\maketitle

The interplay of lattice geometry, nontrivial band topology, and electronic correlations in a kagome lattice could lead to diverse emergent quantum phases of matter \cite{Jiang2023,yin_topological_2022,Neupert2022}.\@ A notable example is the kagome metals $A$V$_3$Sb$_5$ ($A =$ K, Rb, Cs), which has attracted great attention due to the observation of abundant intertwined electronic orders such as nontrivial band topology, CDW, superconductivity, nematicity, and pair density wave \cite{Ortiz2019,Ortiz2020,Ortiz2021,Yin2021,nie_charge_2022,chen_roton_2021}.\@ The recent discovery of CDW order in the magnetic kagome metal FeGe (hexagonal, B35)\@ adds further interest to this field \cite{discoverCDWinFG}:\@ In stark contrast to $A$V$_3$Sb$_5$ and other CDW materials, the CDW state of FeGe appears deep inside the antiferromagnetic phase, suggesting an unconventional CDW mechanism intertwined with magnetism \cite{discoverCDWinFG}.

Despite showing the same $2 \times 2$ charge modulation \cite{discoverCDWinFG,yin_discovery_2022,chen_charge_2023},\@ FeGe and $A$V$_3$Sb$_5$ differ significantly in structural distortions.\@ In $A$V$_3$Sb$_5$, CDW emergence is tied to in-plane distortions within the V-kagome plane and accompanied by phonon instabilities \cite{Tan2021}, likely driven by Fermi surface nesting (FSN) of van Hove singularities (VHSs) \cite{Liu2022_EPC,Xie2022_EPC,Zhong2023_EPC,Kang2022_FSN,Zhou2021_FSN}.\@  Conversely, the CDW in FeGe features dimerization of Ge1 atom along the $c$ axis without showing imaginary phonon modes in the pristine phase \cite{chen_long-ranged_2023,miao_signature_2023,shao_intertwining_2023}.\@ Nonetheless, FSN of kagome states might also drive the CDW transition in FeGe \cite{teng_magnetism_2023,shao_intertwining_2023}.\@  However, the VHS is limited to a narrow $k_z$ window near $E_\mathrm{F}$, and electron-electron correlations might play important roles \cite{wu_electron_2023,ma_theory_2024}.\@  Furthermore, recent photoemission experiments have not shown clear evidence of FSN and gap opening near $E_F$ \cite{zhao_photoemission_2023}.\@ Alternatively, a balance between structural distortion versus magnetic exchange energy, or Peierls-like effects associated with Ge1 bands, can lead to a double-/triple-well energy landscape \cite{wang_enhanced_2023,zhang2023triplewell}.\@ Such a landscape might account for both, the structural distortion and the driving force of the CDW transition. Despite these various proposed mechanisms, the exact nature of the CDW remains elusive so far.  

The exploration of the intrinsic CDW mechanism in FeGe is greatly complicated by discrepancies found experimentally: In early reports, no other transitions were found, except magnetic ones, in the hexagonal phase of FeGe \cite{20susceptibility,20neutron1,20phasediagram,FeGeJPSJ}.\@ In recent neutron and scanning tunneling microscopy (STM) studies, a short-range CDW state with a second-order-like transition was discovered \cite{discoverCDWinFG,yin_discovery_2022,chen_charge_2023}.\@ On the other hand, x-ray scattering experiments indicate that the CDW is long-ranged and its transition is weakly first-order \cite{miao_signature_2023}.\@ The weak CDW signatures and strong sample dependence hamper the investigation of the underlying mechanisms. To settle the experimental discrepancies, and to probe the intrinsic properties, it is of prime importance to prepare high-quality FeGe crystals that show clear and tunable CDW signatures. 

 During the single-crystal growth processes of FeGe, we found that enhancing the CDW signal by varying growth conditions turned out to be challenging. In this Letter, we demonstrate the ability to systematically and repeatably adjust the CDW from short-ranged to long-ranged by employing post-growth annealing in vacuum at temperatures ranging from 240 to \SI{560}{\degreeCelsius}, with subsequent rapid quenching to room temperature in water. The CDW state is characterized by a $2\times2\times2$ superstructure due to the dimerization of Ge1a atoms (see below). High-resolution single-crystal x-ray diffraction (SC-XRD) reveals a significant degree of disorder on the Ge1 sites in the high-$T$ pristine phase of FeGe.\@ This annealing-tunable disorder in the high-$T$ phase holds the key to the fate of both the CDW and various magnetic orders.
 Experimental details can be found in the Supplemental Material (SM) \cite{SM}.

 \nocite{discoverCDWinFG,ishii_post-growth_2021,rotundu_heat_2010,ran_stabilization_2011,sun_review_2019,sayers_correlation_2020,ishida_manifestations_2011,20susceptibility,adams_magnetic_1997,Sheldrick1,Sheldrick2,PetDusPala_2014,huang_fege_2023,arachchige_charge_2022,chen_long-ranged_2023,shao_intertwining_2023,ma_theory_2024,Liu2022_EPC,Xie2022_EPC,Zhong2023_EPC,Kang2022_FSN,Zhou2021_FSN,teng_magnetism_2023,20neutron1,shi_disordered_2023,forsyth_low-temperature_1978,li_spin-reorientation_2018,mandal_spin-reorientation_2011,seo_nearly_2020,xu_magnetic_2017,ludgren_helical_1970,felcher_magnetic_1983,oleszak_structure_2005,kanematsu_magnetic_1965,wu_endotaxial_2018,khalaniya_intricate_2022,verchenko_crystal_2017,grechnev_magnetic_2023,albertini_magnetocrystalline_2004,chen_competing_2024}

\begin{figure}
\includegraphics[scale= 0.34 ]{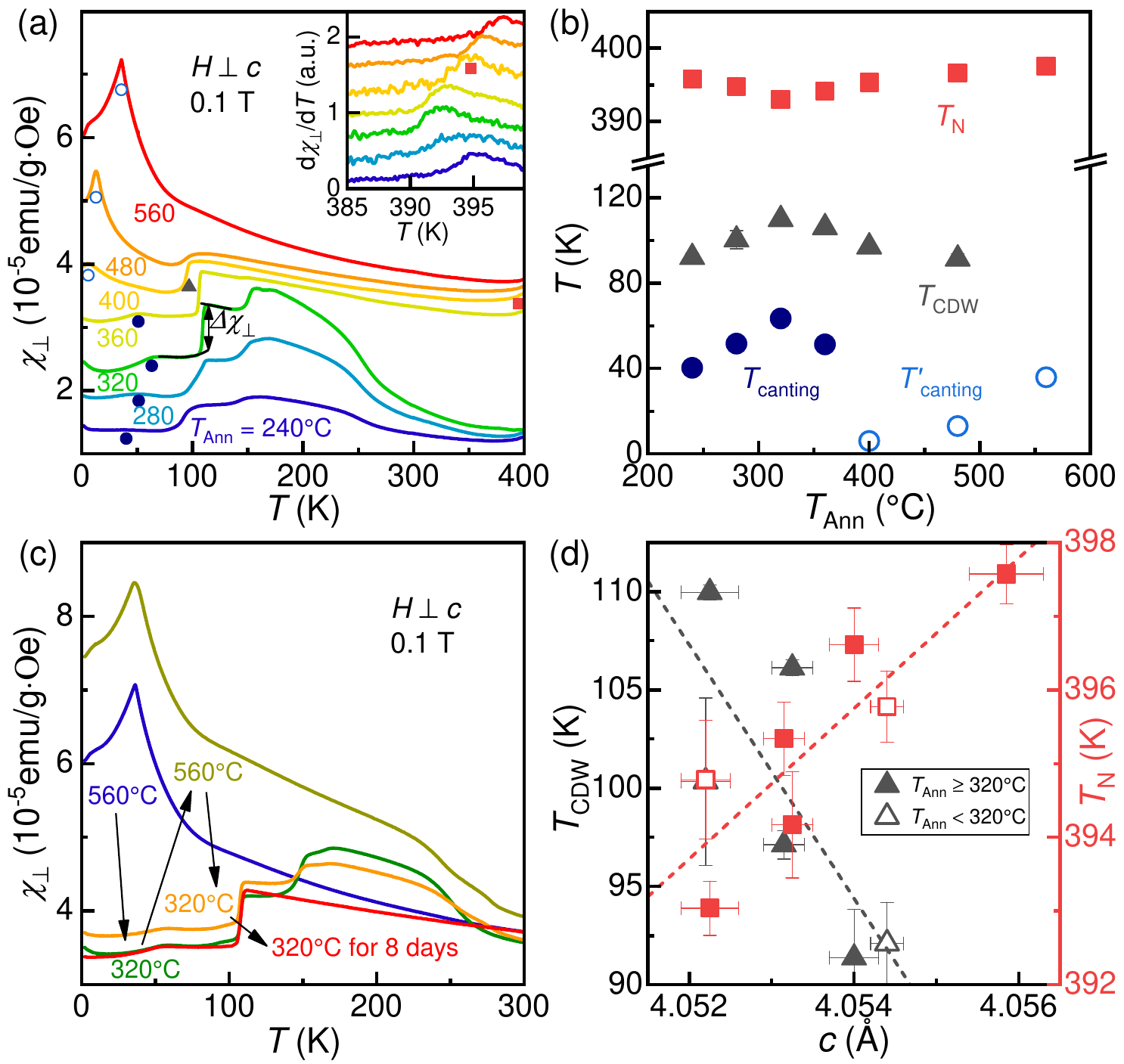}
\caption{(a) Temperature-dependent in-plane magnetic susceptibilities of FeGe samples annealed at different temperatures for 48 h, measured under a 0.1 T magnetic field using zero-field cooling (ZFC) mode. The curves in (a) have been shifted vertically for clarity. The AFM, CDW, and canting transitions are marked by red squares, grey triangles, and blue circles.  The inset in (a) shows the temperature derivative of $\chi_{\perp}(T)$ near $T_\mathrm{N}$. (b) Phase diagram of various transition temperatures tuned by the annealing temperature. (c) $\chi_{\perp}(T)$ of a FeGe crystal annealed in the sequence of 560-320-560-\SI{320}{\degreeCelsius} (48 h each) alongside a sample annealed at \SI{320}{\degreeCelsius} for 8 days (red curve). (d) Correlation of $T_{\mathrm{CDW}}$ and $T_{\mathrm{N}}$ with the $c$-axis lattice parameters. Lattice parameters were obtained from powder x-ray diffraction measurements with a LaB$_6$ standard (Fig. S2 in SM \cite{SM}).\@ Dashed lines are guides to the eye. Note: Error bars in panels (b) and (d) are identical, but are not visible in (b) due to large symbol sizes.}
\label{fig1}
\end{figure}

As shown in Fig. \ref{fig1}(a), the in-plane magnetic susceptibility ($\chi_{\perp}$, $H\perp c$) of FeGe samples varies systematically as a function of annealing temperature, $T_\mathrm{Ann}$.\@ All discrepant behaviors reported in the literature so far can be reproduced by choosing an appropriate $T_\mathrm{Ann}$.\@ Notably, the annealing effect is highly repeatable and reversible. Similar results are obtained for crystals annealed at the same conditions, regardless of their annealing history [see Fig.\@ \ref{fig1}(c)].\@ The overall behavior of the  $T_\mathrm{Ann} = $ \SI{560}{\degreeCelsius} curve is similar to that reported decades ago by Bernhard \emph{et al.} \cite{20neutron1,20susceptibility}: No clear signatures of a CDW transition can be identified. Only the AFM transition (a kink at $T_\mathrm{N} =$ 398 K),\@ and multiple canting transitions (an upturn at $T_{\mathrm{canting}}\sim$ 55 K,\@ a peak at $T'_{\mathrm{canting}}\sim$ 37 K,\@ and a weak kink around $T''_{\mathrm{canting}}\sim$ 6 K),\@ are found.  Early neutron measurements suggest that below $T_\mathrm{N}$, hexagonal FeGe adopts a $c$-axis A-type AFM structure, which transforms to a $c$-axis double cone AFM below $T_{\mathrm{canting}}$ \cite{20neutron1}. Changes in the cone angle produce a peak and a kink in $\chi_{\perp}$ at $T'_{\mathrm{canting}}$ and  $T''_{\mathrm{canting}}$ \cite{20neutron1}.

For $T_\mathrm{Ann}=$ \SI{480}{\degreeCelsius}, a CDW-induced broad jump appears in $\chi_\perp(T)$ around 100 K, similar to that reported by Teng \emph{et al.}\@ \cite{discoverCDWinFG}.\@ With further decreasing $T_\mathrm{Ann}$, the CDW transition becomes much sharper, exhibiting an enhanced transition temperature ($T_\mathrm{CDW}$) and amplitude of the jump in susceptibility ($\Delta\chi_\perp$).\@ For $T_\mathrm{Ann} <$ \SI{320}{\degreeCelsius}, the CDW transition becomes broad again since a longer annealing time would be required if $T_\mathrm{Ann}$ is lowered. The CDW transition is most pronounced for $T_\mathrm{Ann} =$ \SI{320}{\degreeCelsius}, as reflected by the highest $T_\mathrm{CDW}$, sharpest transition width, and the largest $\Delta\chi_\perp$. More importantly, instead of the short-range CDW reported before \cite{discoverCDWinFG,yin_discovery_2022,chen_charge_2023}, a long-range CDW is realized in crystals annealed at \SI{320}{\degreeCelsius}, as evidenced by the SC-XRD shown below in Fig. \ref{fig3}.  

The magnetic orders are also sensitive to the annealing temperatures. The AFM transition temperature $T_\mathrm{N}$ shifts moderately by changing $T_\mathrm{Ann}$ [inset of Fig.\@ \ref{fig1}(a)].\@ With a decrease in $T_\mathrm{Ann}$, the canting transition at $T_{\mathrm{canting}}$ becomes more evident, turning into a step-like feature for $T_\mathrm{Ann} \leq$ \SI{360}{\degreeCelsius}. The transition at $T'_{\mathrm{canting}}$ weakens and shifts below 2 K for $T_\mathrm{Ann} \leq$ \SI{320}{\degreeCelsius}. Note that for $T_\mathrm{Ann} \leq$ \SI{320}{\degreeCelsius}, an additional broad hump appears in $\chi_\perp(T)$ between 150 and 250 K, which can be eliminated by extending the annealing time [red curve in Fig. \ref{fig1}(c)]. Magnetization measurements [Figs. S7-S9 in SM \cite{SM}] suggest that this broad hump is likely induced by a ferromagnetic transition at 250 K and a subsequent spin reorientation at 150 K. This behavior can be ascribed to a small amount ($<$ 0.1\%) of annealing-induced magnetic clusters, which have their easy axis, within the $ab$ plane between 250 and 150 K, rotating to the $c$ axis below 150 K. 

In Fig. \ref{fig1}(b), we summarize the annealing-tuned phase diagram. By varying $T_{\mathrm{Ann}}$, $T_{\mathrm{N}}$ and $T_{\mathrm{CDW}}$ change systematically but with opposite trends: $T_{\mathrm{N}}$/$T_{\mathrm{CDW}}$ reaches its minimum/maximum at $T_\mathrm{Ann}=$ \SI{320}{\degreeCelsius}, indicating an anticorrelation between $T_{\mathrm{N}}$ and $T_{\mathrm{CDW}}$, as seen more clearly in a $T_{\mathrm{N}}$ vs $T_{\mathrm{CDW}}$ plot [Fig. S1(f) in SM \cite{SM}].\@ Crucially, as shown in Fig. \ref{fig1}(d), both $T_{\mathrm{N}}$ and $T_{\mathrm{CDW}}$ show a monotonic variation with the annealing-tuned $c$-axis lattice parameter, indicating a possible structural origin of their anticorrelation. From Fig.\@ \ref{fig1}(b),\@ it can be seen that the evolution of $T_{\mathrm{canting}}$ and $T'_{\mathrm{canting}}$ as a function of $T_\mathrm{Ann}$ are similar to those of $T_\mathrm{CDW}$ and $T_\mathrm{N}$, respectively. The origin of this similarity and the nature of spin-canting transitions will be explored in the following part of the paper.

\begin{figure}
\includegraphics[scale= 0.33]{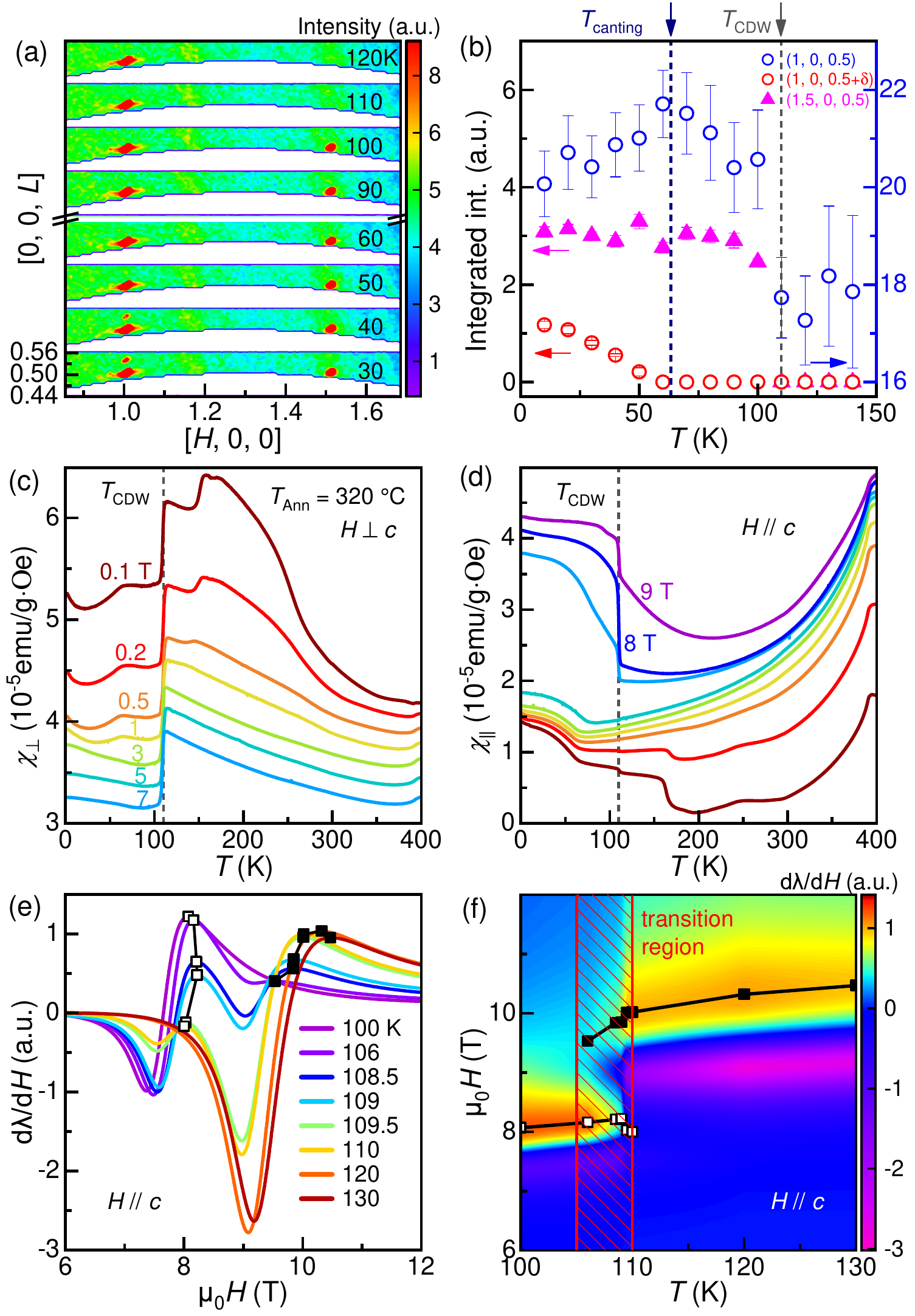}
\caption{ (a) Neutron diffraction pattern of a FeGe crystal annealed at \SI{320}{\degreeCelsius}, captured in the $[H, 0, L]$ plane. (b) Integrated intensities of the AFM peak (1, 0, 0.5), incommensurate magnetic peak (1, 0, 0.5 $+ \delta$), and CDW peak (1.5, 0, 0.5) derived from panel (a) data. Both $y$-axes in panel (b) share the same scale. (c) and (d) shows temperature-dependent in-plane and out-of-plane magnetic susceptibilities, respectively. (e) and (f) Magnetostriction coefficient d$\lambda$/d$H$ and the corresponding $H-T$ phase diagrams, with black squares indicating the spin-flop transition field $H_\mathrm{sf}$. }
\label{fig2}
\end{figure}

To elucidate the nature of the long-range CDW and its relationship with magnetic orders, we conducted single-crystal neutron diffraction, magnetization, and magnetostriction measurements on a \SI{320}{\degreeCelsius} annealed crystal.  As shown in Fig. \ref{fig2}(a), magnetic Bragg peaks associated with the A-type AFM structure are nicely resolved below $T_\mathrm{N}$ at the commensurate magnetic peak (1, 0, 0.5) within the ($H$, 0, $L$) plane \cite{20neutron1,20phasediagram,discoverCDWinFG}.\@ Upon cooling below $T_\mathrm{canting} \sim 60$ K, additional incommensurate magnetic peaks appear at (1, 0, $0.5 \pm \delta$) with $\delta =$ 0.04, signaling a transition to a double cone AFM configuration \cite{20neutron1,20phasediagram,discoverCDWinFG}. An intense peak at (1.5, 0, 0.5) appears just below $T_\mathrm{CDW}=110$ K, indicating CDW formation  \cite{discoverCDWinFG}.\@ The temperature-dependent peak intensities, shown in Fig. \ref{fig2}(b), reveal that the AFM peak intensity sharply increases below $T_\mathrm{CDW}$, implying an intimate correlation between the CDW and magnetic orders \cite{discoverCDWinFG}.\@ Below $T_\mathrm{canting}$, spin canting from the $c$ axis towards the $ab$ plane leads to gradual suppression (enhancement) in the intensity of the commensurate (incommensurate) magnetic peak, agreeing well with previous studies \cite{20neutron1,20phasediagram,discoverCDWinFG}.  The modulation vector of the incommensurate magnetic peak is temperature independent and has almost the same value as that in samples with no CDW \cite{20neutron1} or short-range CDW \cite{discoverCDWinFG,chen_competing_2024}. This result indicates that samples annealed at different temperatures share a similar low-temperature double-cone magnetic structure, with difference in the detailed spin canting transition attributed to variations of cone angles among samples (see Fig.\@ S6 in SM \cite{SM}).\@ Note that the CDW peak intensity shows no anomaly around $T_\mathrm{canting}$, indicating that the CDW is insensitive to the spin-canting transition. 

The low-temperature magnetic structure changes significantly under external magnetic fields. Figs.\@ \ref{fig2}(c) and \ref{fig2}(d) show that the canting transition disappears above 1 T (5 T) for $H \perp c$ ($H \parallel c$). Additionally, the application of a magnetic field above 1 T eliminates the broad hump in susceptibility observed between 150 and 250 K in the low-field conditions, attributed to the saturation of the minor FM clusters. However, the CDW transition remains unaffected by high magnetic fields, even in the spin-flop phase above 7 T for $H \parallel c$.\@ Below $T_{\mathrm{CDW}}$,\@ the susceptibility $\chi_\parallel$ increases sharply above 7 T, reflecting a reduction in the critical field ($H_\mathrm{sf}$) for the spin-flop transition in the CDW state [see Fig. \ref{fig2}(d)].\@ The variation of $H_\mathrm{sf}$ upon the CDW transition is further elucidated through the magnetostriction measurements as depicted in Figs.\@ \ref{fig2}(e) and \ref{fig2}(f).\@ The relative magnetostrictive coefficient d$\lambda$/d$H$ (where $\lambda=\Delta L/L$, and $L$ is the geometrical length of the sample), was measured by a novel composite magnetoelectric method \cite{Chai2021}. Near the CDW transition, d$\lambda$/d$H$ exhibits two dip-peak features associated with the spin-flop transition [Fig. \ref{fig2}(e)].\@ On the other hand, only a single dip-peak feature is evident in d$\lambda$/d$H$ both below and above $T_\mathrm{CDW}$.\@  As summarized in Fig.\@ \ref{fig2}(f),\@ $H_\mathrm{sf}$ exhibits a sharp step-like jump, accompanied by a phase coexistence region near $T_\mathrm{CDW}$, indicating the coexistence of high-temperature pristine and low-temperature CDW phases.  This serves as compelling evidence for the first-order nature of the CDW transition. These findings suggest that the CDW is unaffected by changes in the magnetic structure, while $H_\mathrm{sf}$ is markedly reduced in the CDW phase.

\begin{figure*}
\includegraphics[scale= 0.6]{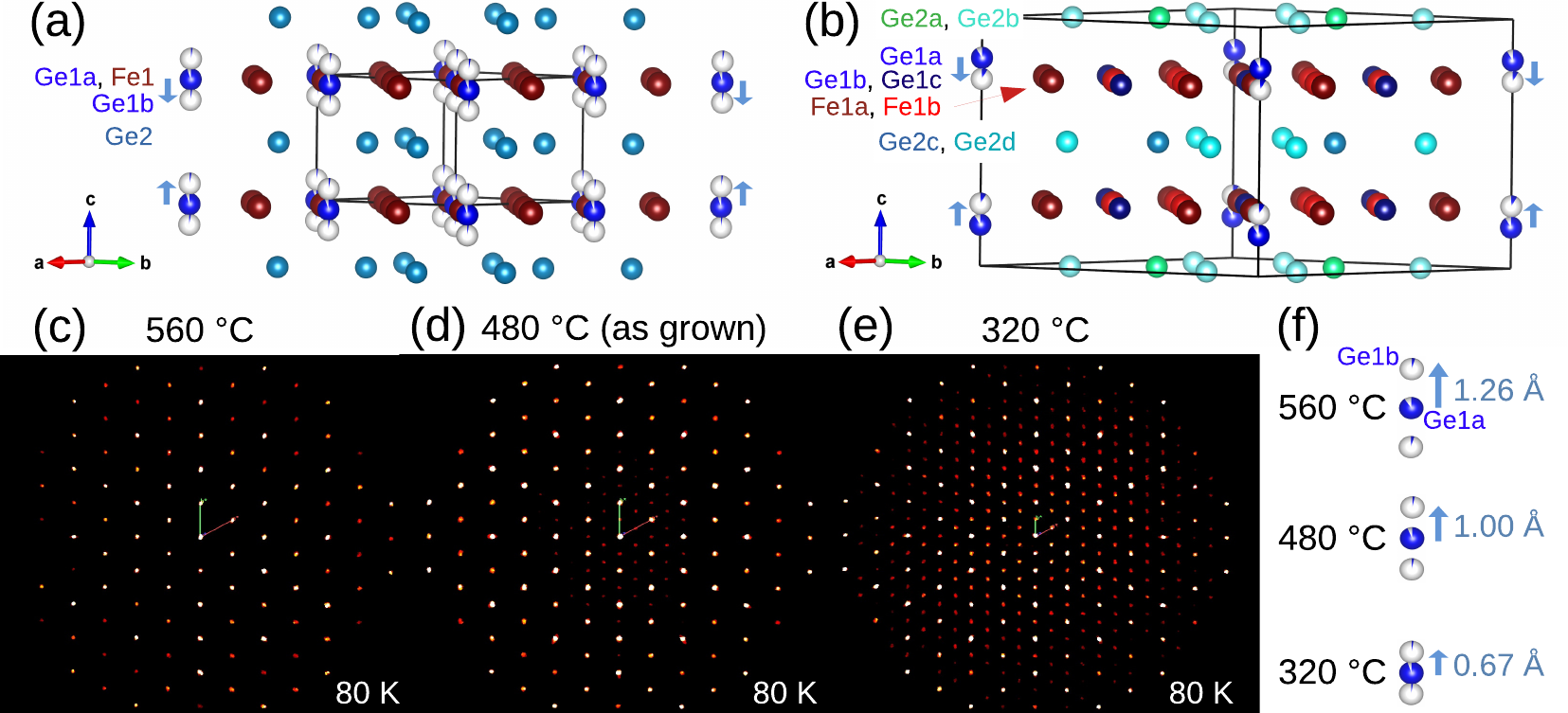}
\caption{(a) Crystal structure of pristine FeGe at 300 K.\@ In addition to the established Ge1a (the Ge site in the Fe$_3$Ge plane), a disordered Ge1b site is found shifted along the crystallographic \textit{c} axis as indicated by the light blue arrows. The degree of disorder strongly depends on the annealing conditions and is reflected in the occupancy of the disordered site and its distance to Ge1a as shown in (f).\@ (b) $2 \times 2 \times 2$ CDW crystal structure of FeGe at 80 K realized only for samples annealed at \SI{320}{\degreeCelsius}.\@ To form the Ge1a-Ge1a dimers, the Ge1a atoms are moved upwards (downwards) from their initial positions. However, also for the low-$T$ structure a disordered Ge1b site is found at the position where the Ge1a atoms initially reside at 300 K.\@ (c), (d), and (e) illustrate the RL view along the $c^{*}$ direction measured at 80 K for samples annealed at  \SI{560}{\degreeCelsius},\@ \SI{480}{\degreeCelsius}, and \SI{320}{\degreeCelsius},\@ respectively. The high-$T$ RL of all investigated samples %(presented in the SM) 
looks essentially identical to the one in (c).\@ (f) Distance at 300 K between Ge1b and Ge1a which is a reasonable indicator for the amount of disorder. }  
\label{fig3}
\end{figure*}

Finally, our SC-XRD results are displayed in Fig.\@ \ref{fig3}.\@ Analysis of the 300 K crystallographic structure in space group $P6/mmm$ initially suggests a replication of known atomic positions. However, profound refinement of all samples indicates incomplete occupancy of the Ge1a site, i.~e.,\@ of the Ge site within the Fe$_3$Ge plane. Difference Fourier analysis further reveals a significant residual electron density shift along the $c$ axis away from the Ge1a site, as detailed in the SM \cite{SM}.\@ Inserting a Ge1b atom at this position, the discrepancy in Ge occupancy at the Ge1a site is precisely balanced by the Ge1b site (with and without restricting the sum of both sites to an occupation of 1), markedly improving refinement residuals, as outlined in the SM \cite{SM}. Fig. \ref{fig3}(a) illustrates the partially occupied Ge1a and the disordered Ge1b site. This finding unambiguously demonstrates that the high-$T$ phase of FeGe features substantial disorder. 

Most interestingly, the occupancy of the disordered Ge1b site and its distance to the Fe$_3$Ge plane strongly depend on the annealing conditions of the samples as illustrated in Fig.\ \ref{fig3}(f) and serve as a good indicator for the amount of disorder: Samples annealed at \SI{560}{\degreeCelsius},\@ \SI{480}{\degreeCelsius},\@ and \SI{320}{\degreeCelsius} show occupancy of the disordered Ge1b site (of Ge1a) of 9 \% (91 \%),\@ 6 \% (94 \%),\@ and 5 \% (95 \%) and a distance to Ge1a of 1.259(9) {\AA},\@ 1.00(2) {\AA},\@ and 0.665(9) {\AA},\@ respectively. Hence, with a higher annealing temperature, the disorder determined at 300 K strongly increases. Since SC-XRD averages over many unit cells, it is difficult to decide whether individual disordered Ge1b sites are statistically distributed over the sample or if they randomly occur in pairs in a unit cell. If in pairs, however, they could be interpreted as preformed Ge1b-Ge1b dimers locally existing already far above $T_{\rm CDW}$ which would be consistent with what is observed for FeGe$_{1-x}$Sb$_x$ in Ref.\@ \cite{huang_fege_2023}.\@

The 80 K reciprocal lattice (RL) views along $c^{*}$ plotted in Figs.\@ \ref{fig3}(c),\@ (d),\@ and (e) exemplify that a long-range CDW and, thus, the low-$T$ $2 \times 2 \times 2$ superstructure of FeGe shown in Fig.\ \ref{fig3}(b) is realized only for samples annealed at \SI{320}{\degreeCelsius}.\@ For samples annealed at \SI{480}{\degreeCelsius},\@ only weak and more diffuse reflections characteristic for a short-range CDW are detected, whereas for samples annealed at \SI{560}{\degreeCelsius} CDW reflections are completely absent.\@ These observations indicate that the CDW volume fraction decreases with increasing $T_\mathrm{Ann}$, consistent with the magnetization measurements (see Fig.\@ \ref{fig1}).\@ For samples annealed at \SI{320}{\degreeCelsius},\@ the Ge1a atoms are moved upwards (downwards) from their initial positions at 300 K to form a $2 \times 2$ pattern of Ge1a-Ge1a dimers [Fig. \ref{fig3}(b)].\@ Remarkably, a disordered Ge1b position (occupation 0.08) shifted from the low-$T$ Ge1a dimer position  (occupation 0.92) is also found in this phase [Fig.\ \ref{fig3}(b)].\@ Together with the formation of Ge1a-Ge1a dimers, small but clear in-plane Kekul\'{e}-type distortions in the two Ge2 honeycomb layers and predominant out-of-plane buckling in the two Fe kagome layers can be observed (see SM \cite{SM} for details).\@  Apart from tiny in-plane movements of the Fe atoms resolved in our data \cite{SM},\@ these findings are consistent with theoretical predictions \cite{shao_intertwining_2023,zhou_magnetic_2023,wang_enhanced_2023}.\@ In contrast, the low-$T$ structures of samples annealed at \SI{480}{\degreeCelsius} and \SI{560}{\degreeCelsius} are similar to their high-$T$ phase with subtle variations in the Ge1b disorder [see Fig. \ref{fig3}(a)] resulting in an identical occupation of the Ge1b site (8.5 \%  for both samples) and Ge1b-Ge1b distances between 2.46$-$2.69 {\AA}.\@ These low-$T$ Ge1b-Ge1b bond lengths are close to the Ge1a-Ge1a dimer distance (2.72 {\AA}) of \SI{320}{\degreeCelsius} annealed samples and, therefore, perfectly fit into a dimer picture.\@ Accordingly, a sound rationale for the variance in the CDW volume fraction lies in the distribution of the Ge1 dimers:\@ Statistically for the \SI{560}{\degreeCelsius} sample, clustered for the short-range CDW ordered \SI{480}{\degreeCelsius} sample, and long-range $2 \times 2$ ordered for the genuine CDW phase of the \SI{320}{\degreeCelsius} sample. A more detailed and complementing crystallographic discussion of all structural properties and atomic positions will be given in the SM \cite{SM}.

Our SC-XRD analysis thus reveals multiple energy minima for Ge1 atoms, characterized by specific distances ($z$) to the Fe$_3$Ge plane: Pristine Ge1a at $z =$ 0, a range of positions for Ge1b with $z = 0.16-0.3$, and the Ge1a-Ge1a dimer at $z=$ 0.16. These findings are consistent with predictions from double/triple-well potential models \cite{wang_enhanced_2023,zhang2023triplewell}, which might be best captured in the framework of a multi-well potential model. The distribution of Ge1 atoms at 300 K is largely determined by the annealing conditions:\@ Higher annealing temperatures increase the probability of Ge1 atoms populating metastable Ge1b sites, whereas lower annealing temperatures lead to the occupation of the slightly more stable pristine Ge1a sites. Consequently, high Ge1b disorder levels trap the system in local energy minima, hindering the transition to the possible lower-energy CDW state with $2 \times 2$ Ge1a-Ge1a dimers at low temperatures.\@ Varying annealing conditions, however, can overcome the energy barrier between energy minima, allowing for repeatable and reversible ground state tuning, as evidenced in Fig.\@ \ref{fig1}(c).\@

The annealing-tuned magnetic properties are also likely linked to the displacement of Ge1 atoms, significantly influencing the magnetic exchange interaction ($J$), anisotropy ($K$), and spin splitting/polarization. The displacement of Ge1b atoms away from the Fe$_3$Ge plane leads to a more three-dimensional (3D) structure, potentially enhancing the interlayer $J_c$ and reducing $K$.\@ This is evidenced by the increase of $T_\mathrm{N}$ with higher annealing temperatures [Fig.\@ \ref{fig1}(b)], mirroring trends observed in van der Waals layered magnets \cite{seo_nearly_2020,yan_type_2020}.\@ Consequently, the observed anticorrelation between $T_\mathrm{N}$ and $T_\mathrm{CDW}$ may stem from their distinct relationships with Ge1b disorder. The increased magnetic moment (decreased $H_\mathrm{sf}$) in the CDW phase is most probably due to Ge1-distortion-induced enhancement of spin splitting/polarization (suppression of magnetic anisotropy) \cite{wang_enhanced_2023,shao_intertwining_2023}. Furthermore, Ge1 disorder, which locally breaks the inversion symmetry, could induce finite Dzyaloshinskii–Moriya interaction.\@ This offers a natural explanation for the observed symmetry-forbidden double cone magnetic structure at low temperatures and the persistence of incommensurate spin excitations up to at least 350 K \cite{zhou_magnetic_2023,chen_competing_2024}. The spin canting transition is thus tied with the Ge1 displacements. In samples dominated by the CDW phase ($T_\mathrm{Ann} \leq$ \SI{360}{\degreeCelsius}),  $T_\mathrm{canting}$ exhibits a similar $T_\mathrm{Ann}$ dependency as $T_\mathrm{CDW}$,\@ while it resembles the one of $T_\mathrm{N}$ in samples dominated by Ge1b disorder ($T_\mathrm{Ann} \geq$ \SI{400}{\degreeCelsius}). These findings suggest that structural factors play a dominant role in FeGe, with magnetism possibly being a secondary effect.\@ This is similar to the structurally dominated CDW observed in non-magnetic materials such as ScV$_6$Sn$_6$ and IrTe$_2$ \cite{lee_nature_2024,tan_abundant_2023,kim_origin_2015}.

In conclusion, our study demonstrates that annealing treatments can reversibly tune both the CDW and magnetic orders in FeGe. The annealing-induced variation of CDW and magnetic orders, as well as their relationship, are closely linked to Ge1 atom displacement, which is adjustable through annealing. This study offers a notable method for achieving a long-range CDW in FeGe, emphasizes the predominant influence of structural factors, and sets important constraints on the theoretical understanding of the CDW mechanism.

\begin{acknowledgments}
We thank Ya-Jun Yan,\@ Yilin Wang, Yuan Li,\@ Christoph Meingast,\@ Matthieu Le Tacon,\@ and Weijun Ren for their helpful discussions. This work was supported by the Natural Science Foundation of China (No.\@ 12374081,\@ No.\@ 12227806,\@ and No.\@ 12004056.)\@ A.W. acknowledges the support from Chongqing Research Program of Basic Research and Frontier Technology, China (Grants No. cstc2021jcyj-msxmX0661).\@ Y.S.C. acknowledges the support from the Beijing National Laboratory for Condensed Matter Physics. M. He acknowledges the support by Chinesisch-Deutsche Mobilit\"atsprogamm of Chinesisch-Deutsche Zentrum f\"ur Wissenschaftsf\"orderung (Grant No. M-0496).\@ We would like to thank Guiwen Wang and Yan Liu at the Analytical and Testing Center of Chongqing University and Siegmar Roth and Andre Beck at the Institute for Quantum Materials and Technologies,\@
Karlsruhe Institute of Technology for their technical assistance.
\end{acknowledgments}

%\bibliographystyle{apsrev4-2}
%\bibliography{FeGe}% Produces the bibliography via BibTeX.

\end{document}